\documentclass[a4paper]{article}
\usepackage[a4paper, left=3.0cm, right=3.0cm, top=2cm, bottom=2cm]{geometry}

\usepackage{url}
\usepackage[latin1]{inputenc}
\usepackage{amsmath}

\usepackage{graphicx}
\usepackage{xcolor}
\usepackage[urlcolor=blue]{hyperref}      
 \hypersetup{
     colorlinks = true,                    
     citecolor = {blue},
     linkcolor = {purple},
            }

\usepackage{natbib}
\setcitestyle{authoryear,open={(},close={)}}  
\setcitestyle{aysep={,}}                      
\setcitestyle{semicolon}                      

\usepackage{gensymb}       

\usepackage{hyperref}

\usepackage{booktabs}

\usepackage{colortbl}
\usepackage{pstricks,pst-grad}
\usepackage{pst-pdf}

\usepackage{fancybox, graphicx}
\graphicspath{{./}{./figs/}}

\definecolor{lightBrown}{rgb}{0.9,0.8,0.6}
\definecolor{lightGreen}{rgb}{0.8,1.0,0.8}
\definecolor{lightYellow}{rgb}{1.0,1.0,0.8}
\definecolor{lightBlue}{rgb}{0.0,0.4,0.8}
\definecolor{lightRed}{rgb}{1.0,0.8,0.8}
\definecolor{darkYellow}{rgb}{0.9,0.7,0.1}
\definecolor{titleBlue}{rgb}{0.0,0.2,0.6}

\newrgbcolor{white}{1.0 1.0 1.0}
\newrgbcolor{lightGreen}{0.8 1.0 0.8}
\newrgbcolor{lightYellow}{1.0 1.0 0.8}
\newrgbcolor{lightGrey}{0.6 0.6 0.6}
\newrgbcolor{lightBlue}{0.0 0.4 0.8}
\newrgbcolor{mediumRed}{0.4 0.1 0.1}
\newrgbcolor{darkCyan}{0.25 0.5 0.5}
\newrgbcolor{darkGrey}{0.4 0.4 0.4}
\newrgbcolor{darkBrown}{0.3 0.2 0.1}
\newrgbcolor{darkBlue}{0.0 0.1 0.3}
\newrgbcolor{darkGreen}{0.1 0.3 0.1}
\newrgbcolor{darkOrange}{0.8 0.4 0.0}
\newrgbcolor{darkRed}{0.6 0.2 0.2}
\newrgbcolor{darkViolet}{0.2 0.0 0.2}
\newrgbcolor{darkYellow}{0.9 0.7 0.1}
\newrgbcolor{veryDarkCyan}{0.20 0.4 0.4}

\newlength{\parboxWidth}          

\begin{document}
\title{\bf Charting closed-loop collective cultural decisions:
From book best sellers and music downloads to Twitter 
hashtags and Reddit comments}
\author{Lukas Schneider, Johannes Scholten, Bulcs\'u S\'andor$^*$, Claudius Gros 
\\
Institute for Theoretical Physics, Goethe
University Frankfurt, Germany \\
$^*$Department of Physics, Babes-Bolyai University, Cluj-Napoca, Romania
} 


\maketitle
\begin{abstract}
Charts are used to measure relative success for 
a large variety of cultural items. Traditional 
music charts have been shown to follow self-organizing
principles with regard to the distribution of 
item lifetimes, the on-chart residence times. 
Here we examine if this observation holds also 
for (a) music streaming charts (b) book best-seller 
lists and (c) for social network activity charts, 
such as Twitter hashtags and the number of comments 
Reddit postings receive. We find that charts based 
on the active production of items, like commenting, 
are more likely to be influenced by external factors, 
in particular by the 24\,hour day-night cycle. External 
factors are less important for consumption-based 
charts (sales, downloads), which can be explained by a
generic theory of decision-making. In this view, 
humans aim to optimize the information content of 
the internal representation of the outside world, 
which is logarithmically compressed. Further support 
for information maximization is argued to arise 
from the comparison of hourly, daily and weekly charts, 
which allow to gauge the importance of decision times 
with respect to the chart compilation period.  
\end{abstract}


\large
\section{Introduction}

Isaac Asimov's science fiction `Foundation' trilogy 
is based on the presumption that it may be 
possible to develop a framework which would allow 
to model and predict the political and economic 
development of human societies as such 
\citep{elkins1976isaac,phillips1999learning}. At
the core of the framework in question, `psychohistory', 
lies the hypothesis that it could be possible to
describe collective decision-making by rigorous laws,
regardless of what a specific individual decides 
to do. Individual preferences would average out when 
populations are large. On a decisively less galactic 
level, a legitimate question regards the statistics 
of cultural phenomena produced by large numbers of 
decisions. Prime candidates are in this regard rankings
measuring the economic success of cultural goods, 
like music albums, books and movies. Here we will 
show that certain traits of the New York Times best-seller 
list and the streaming charts compiled on a daily basis by Spotify,
a streaming portal, are congruent with the predictions of
an information-theoretical theory of human decision-making.

It has been recognized that charts, in particular 
music charts, like the US-based Billboard charts, 
provide access to the study of an extended range of 
socio-cultural developments. Examples are the evolution 
of harmonic and timbral properties of popular music 
\citep{mauch2015evolution}, the influence of gender 
and race on chart success \citep{lafrance2018race}, 
and how technological progress influences cultural 
diversity, respectively concentration processes 
\citep{ordanini2016fewer}. Other studies investigated 
the correlation between acoustic features
\citep{interiano2018musical} and repetitive lyrics
\citep{nunes2015power} to market success, the interplay
between popularity and significance in popular music
\citep{monechi2017significance}, and whether there is a 
trend towards a global musical culture 
\citep{achterberg2011cultural}. An alternative venue
for the study of long-term societal developments are 
recognized scientific and cultural awards, like the 
Nobel price \citep{gros2018empirical}.

On an extended level, \citet{schneider2019five} suggested 
that music charts are suitable proxies for testing
quantitatively the concept of social acceleration
\citep{rosa2013social}, viz the notion that both 
personal and societal processes proceed nowadays faster 
than they used to \citep{wajcman2016sociology,vostal2017slowing}.
The rationale for this suggestion is based on the observation
that long-term databases of comparable entries are indispensable
for studies aiming to measure the evolution of culturally and 
sociologically relevant time scales, a condition satisfied by 
charts, music and book charts alike, which provide timelines 
of five and more decades of weekly compiled data. For the
US, UK, German and Dutch music charts it was found
\citep{schneider2019five}, that the evolution of the album
lifetimes and of the weekly rank decay provides evidence for 
a substantial acceleration, in particular since the rise of 
the internet. These findings are congruent with the changes
observed for the chart diversity and for the dynamics of 
number-one hits.

Cultural items, like books and music albums, compete for
resources, such as monetary budgets, attention and dedication.
Competition for time and attention is also a driver
for the dynamics of news topics on social networks, a 
process that has been argued to take place on accelerating 
time scales \citep{lorenz2019accelerating}. The success of 
social media generated content is closely connected to 
sharing, which is an aspect not present for classical 
cultural items. It has been observed in this context,
that the statistics of sharing and scientific citations 
seem to follow identical universal 
functionalities \citep{neda2017science}.

An accessible example of a social platform is Reddit
\citep{medvedev2017anatomy}, a user generated exchange 
platform for information, opinions and comments. 
Studies of Reddit data examined \citep{baumgartner2020pushshift},
among other things, the relation between popularity and
intrinsic quality \citep{stoddard2015popularity},
the stability of online communities with substantial
turnover rates \citep{panek2018effects}, and the
evolution of users interests \citep{valensise2019drifts}.

Taking the number of comments as a popularity
measure, one can generate charts which parallel
download and sales charts. One key difference is
however that the exact timing of the triggering
event is now of relevance. Download charts
contain albums that have been released not only
the same day, but also considerably in the past.
New posts remain in contrast popular on Reddit only
for a maximum of one or two days, which implies
that the day-night activity cycle of the user
could impact the lifetime distribution of Reddit 
posts. This is indeed what one finds. We argue
that this observation implies that the feedback 
between popularity rankings and user activity 
interferes with additional time scale, like the 
24\,hour day-night activity cycle, when
the timing of the user activity is strongly
correlated with that of the triggering event.

Book and music sales and streaming charts are 
primarily consumption charts, for which an 
aggregate information-theoretical theory is
proposed. Exactly when a consumption takes place,
like the acquisition or the downloading of books 
and music, is not relevant, in contrast to the posting
of Reddit comments, and other social media activities, 
which are time-relevant because they involve the 
creation of new items. As a consequence, we find
when analyzing Reddit and Twitter comment and sharing 
charts, that additional driving forces, in particular
the 24h day-night cycle, tend to mask the development 
of fully autonomous self-organization for charts of
social media activities.





\section{Information theory of human decision-making}

On a statistical level, human decision-making may be 
modeled using an information theoretical ansatz
\citep{gros2012neuropsychological}. Starting from
the same basic ansatz as \citet{schneider2019five}, 
we present here a modified derivation. From
an epistemology view, we stress that our ansatz 
is intended as a possible explanatory framework.
Similarly, the data analysis presented further
below is not claimed to proof the correctness of 
the information theoretical ansatz, indicating however that
it is a valid contender. From a generalized 
perspective our ansatz is neutral 
\citep{leroi2020neutral}, in the sense that is
based not on semantic, but on statistical 
considerations.

\subsection{Decision-making maximizing information\label{sect_H_max}}

The brain processes and stores information. Given
the finite size, information needs to be selected
and compressed \citep{marois2005capacity}.
A suitable measure for the amount of information 
encoded by the probability distribution $p(s)$ of 
a quantity $s$ is given by the Shannon Entropy 
$H[p]$,
\begin{equation}
H[p] = -\langle \log p\rangle,
\qquad\quad
\langle A\rangle = \int ds\, A(s)\,p(s)\,,
\label{H_p}
\end{equation}
where $\langle A\rangle$ denotes the expectation
value of the function $A=A(s)$, see e.g.\
\citet{gros2015complex}. We assume that the 
overall goal of human decisions is to maximize
information, in particular the information 
generated by decision-making processes. As 
a proxy for the impact of actions, e.g.\ to 
buy a cultural item, we examine the 
statistical properties of the corresponding 
chart.

Statistically, information maximization is 
equivalent to maximizing the entropy $H[p]$.
This goal can be achieved in practice only 
when respecting a given set of constraints. 
Real-world actions typically need to factor 
in the amount of effort and time involved,
as well as the uncertainty of the outcome, viz
the variance. On a statistical level these two
constraints are given by the mean and the 
variance, which are equivalent to the first 
and the second moment of the probability 
distribution in question, $\langle s\rangle$ 
and $\langle s^2\rangle$. The constrained 
maximum entropy distribution is consequently obtained
by maximizing the objective function
\begin{equation}
\Phi[p] = H[p] - a \langle s\rangle - b \langle s^2\rangle\,,
\label{Phi_p}
\end{equation}
where $a$ and $b$ are suitable Lagrange multipliers. Using 
standard variational calculus \citep{gros2015complex},
the distribution function maximizing $\Phi[p]$ is found 
to be a Gaussian with mean $\mu=-a/(2b)$ and variance 
$\sigma^2=1/(2b)$,
\begin{equation}
p(s) \sim \mathrm{e}^{-a s-b s^2}
\sim \mathrm{e}^{-(s-\mu)^2/(2\sigma^2)}\,,
\label{p_s_H_max}
\end{equation}
where normalization factors have been
suppressed.

\begin{table}[b!]
\centering
\caption{{\bf Parameters of lifetime distributions.} Listed are 
the linear and quadratic parameters $a$ and $b$ from Eq.~\eqref{p_L_H_max}. 
Note that $a+1$ corresponds to the exponent of the power law 
contribution, see (\ref{p_L_power_law}). The corresponding 
quadratic curves in log-log space are shown in 
Figure~\ref{lifetime_distribution_music}
and~\ref{fig_SpotifySingle_TwitterHastags}. 
The table is sorted with respect to the linear 
parameter $a$. The uncertainties refer to the margin of
error given a 95\% confidence level. 
}
\label{tab_lifetime_distribution_parameters}
\begin{tabular}{lrr}
\\
\toprule
chart & $a\!+\!1$ & $b$\\
\midrule
Billboard Album (early) & $ -2.21 \pm 0.45$ & $ 1.5 \pm 0.22$\\
Spotify Album daily & $ 1.43 \pm 0.23$ & $ 0.89 \pm 0.40$\\
Twitter hourly 1 & $ 1.43 \pm 0.75$ & $ 0.27 \pm 0.45$\\
Spotify Single Weekly & $ 1.21 \pm 0.20$ & $ 0.11 \pm 0.10$\\
Billboard Album (late) & $ 1.79 \pm 0.45$ & $ -0.1 \pm 0.29$\\
Spotify Single Daily & $ 1.68 \pm 0.21$ & $ -0.12 \pm 0.07$\\
Spotify Album weekly & $ 2.54 \pm 0.75$ & $ -0.12 \pm 0.74$\\
Twitter daily & $ 3.16 \pm 0.28$ & $ -0.56 \pm 0.11$\\
Twitter weekly & $ 3.05 \pm 0.70$ & $ -0.68 \pm 0.38$\\
Twitter hourly 2 & $ 6.63 \pm 1.36$ & $ -1.18 \pm 0.35$\\
\bottomrule
\end{tabular}
\end{table}

\subsection{Human brains compress information 
            logarithmically\label{sect_Weber_Fechner}}

When making decisions, which is the distribution 
function $p(s)$ that is to be maximized statistically
using the entropy $H[p]$? For the case of
book and music charts, a prime candidate is the lifetime
distribution $p(L)$, which measures the probability
that a given item, a book or a music album, remains 
listed for a period $L$, the lifetime. While listed,
books and albums receive increased attention, which
implies that the listing period constitutes a visible 
effect of the individual decision to buy a specific 
cultural item.

From the cognitive perspective, the brain is confronted with 
two contrasting demands: to store incoming information 
as faithfully as possible, covering at the same time 
the extended orders of magnitude characterizing physical 
stimuli, like sound and light intensity, as well as 
time scales. An efficient solution for this conundrum
is to store information on a compressed scale,
for instance logarithmically. This is indeed the case, as
expressed by the Weber-Fechner law, which states
that the brain discounts sensory stimuli
\citep{hecht1924visual}, numbers \citep{nieder2003coding,dehaene2003neural}
and time logarithmically \citep{howard2018memory}.
It is not a coincidence that we use logarithmic
scales, lumen and decibel, to measure light and
sound intensities.

\subsection{Maximizing compressed information\label{sect_infoMax}}

When performing an operation, like information 
maximization, the brain uses internal states that 
we denote here $s$. These states are in general
related logarithmically to outside quantities,
as discussed in the previous section. For the case 
of the lifetime distribution $p(L)$ we have 
\begin{equation}
s=\log(L), \qquad\quad
p(s)ds = p(L)dL\,,
\label{s_log_L}
\end{equation}
which relates the observable distribution function
$p(L)$ with its internal representation, the probability 
density $p(s)$.  Using $ds/dL=1/L$ and the maximum 
entropy probability density $p(s)$, as given
by (\ref{p_s_H_max}), one finds
\begin{equation}
p(L) \sim \frac{1}{L}\mathrm{e}^{-a \log(L)-b \log^2(L)}
\sim \mathrm{e}^{-(\log(L)-\tilde\mu)^2/(2\tilde\sigma^2)}\,,
\label{p_L_H_max}
\end{equation}
with a log-mean $\tilde\mu=-(a\!+\!1)/(2b)$, and 
a log-variance of $\tilde\sigma^2=1/(2b)$.
$p(L)$ corresponds to a log-normal distribution
whenever $b>0$, and to a power law,
\begin{equation}
p(L) \sim \frac{1}{(L)^{a+1}},
\qquad\quad b\to0\,,
\label{p_L_power_law}
\end{equation}
if $b$ vanishes. Power laws, which are frequently 
observed \citep{markovic2014power}, are hence a natural
outcome of human activities, statistically, whenever
the variance is either small or not taken into account. 

There are two causalities for why the variance may
be of secondary importance. Firstly, when uncertainties
are small, viz when fluctuations around the mean are
negligible. In this case the $b$-term in (\ref{p_L_H_max}) 
is numerically small, becoming important only for 
exceedingly large or small lifetimes $L$. 
A second, more general argument is that one
has to invest comparatively more resources 
to sample the variance than just the mean,
in particular when correlations are present 
\citep{broersen1998estimation}. If time is a
scarce resource, only the mean can be obtained reliably 
from a time dependent functionality, such as the 
lifetime distribution of cultural items. As a
consequence it follows, that power laws are present 
when individual decision times are shorter than typical 
chart listing duration, with log-normal distributions 
emerging for extended decision times.

The probability density $p(L)$ is normalizable only
for $b\ge0$ if arbitrary large lifetimes $L$ are 
allowed. Real-life time intervals are however finite,
which makes a negative $b$ also viable. The values for
$a+1$ and $b$ obtained throughout this study for different 
music and social network activity charts are discussed in the 
next section. 

\begin{figure}[t]
\centerline{
\includegraphics[width=1.00\columnwidth]{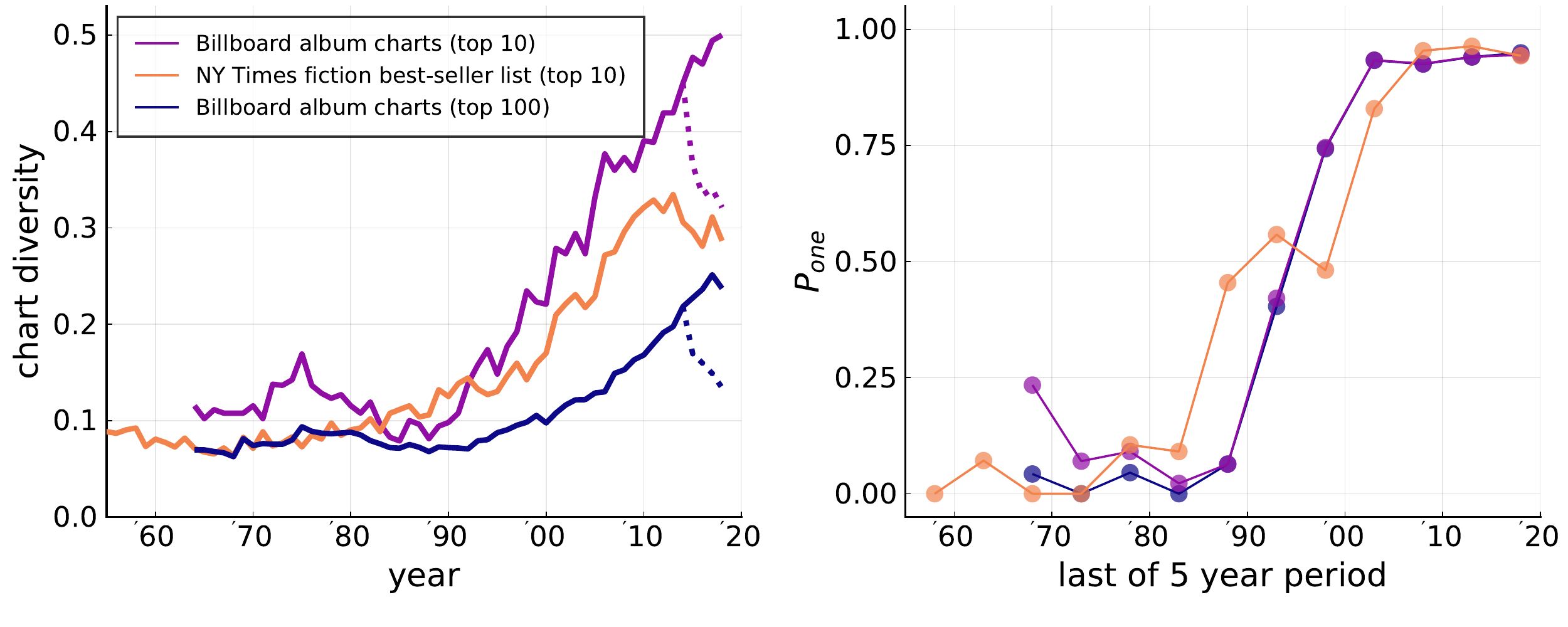}
           }
\caption{{\bf Long-term evolution of charts and best-seller lists.} 
{\em Left:} The chart diversity $d=N_a/N_s$, with $N_a$ denoting
the number of unique titles  on a chart with $N_s$ slots. The 
chart diversity of the New York Times best-seller list (orange) 
follows a trend similar to the top 10 of the Billboard 
album chart (purple). The top 100 album chart (blue) is 
also shown for comparison. The dotted lines indicate the 
inclusion of streaming data into the Billboard album chart
(see Sect.~\ref{sect_charts} for details).
{\em Right:} Out of all number-one best sellers/hits, the 
fraction $P_{\rm one}$  of books/albums that top the
best-seller list/charts already on entry, viz the first 
time they are listed. Initial success is all important
for today's cultural items.
}
\label{nytimes_cd_pone}
\end{figure}

\section{Results}

The information-theoretical approach to 
the statistics of human decision-making 
implies a characteristic functional dependency 
for the distribution of the lifetime
of cultural items. We tested this dependency
for book and album sales charts, for music
download charts, as well as for social media 
comments. Data retrieval and processing is
described in Sect.~\ref{sect_data}. To visualize 
the possible $p(L)$ distribution functions,
figures with log-log representations of lifetime 
distributions include quadratic fits that correspond to
Eq.~\eqref{p_s_H_max}. However, these fits are not
claimed to correspond to statistically
validated models, as discussed, e.g.,
by \cite{clauset2009power}. In 
Table~\ref{tab_lifetime_distribution_parameters} 
and~\ref{tab_nytimes_lifetime_distribution_parameters} 
the respective parameters are listed.

\subsection{Chart diversity \label{sect_nytimes_chart_diversity}}

An intuitive measure for the turnover rate of charts
is the chart diversity $d = N_a/N_s$, where $N_a$ is 
the number of unique titles, here per year, normalized by
the number of available slots $N_s$ (for the whole year). 
Figure~\ref{nytimes_cd_pone} shows the evolution of the 
chart diversity of the New York Times best-seller list,
in comparison with the top 10 and top 100 Billboard 
album charts. Billboard introduced streaming data into 
the chart in 2014 but still published a sales chart. 
This is why the line splits in 2014, with 
the chart including streaming shown as a dotted line.

The chart diversities of the New York Times best-seller 
list and of the Billboard album charts have taken a similar
evolution over the course of the last three to four decades. 
This fact is quite remarkable, given that individual music 
albums and fiction books differ substantially concerning
their respective consumption times. Generally one
needs substantially longer time to read a novel than to listen 
to an album. Their overall key core characteristics are
on the other hand similar. The average number 
of weeks a novel spent on the New York Times best-seller
list $\bar{w}\approx 1/d$ (where $d$ is the chart diversity),
shrunk from $\bar{w} \approx 1/0.1 = 10$ weeks in the 1980s
to about $\bar{w} \approx 1/0.3 \approx 3 $ in 2010. The 
equivalent
numbers for the top-10 Billboard album charts are
$\bar{w} \approx 1/0.1 = 10$ in the 1980s and 
$\bar{w} \approx 1 /0.4 = 2.5$ in 2010.

\begin{table}[b!]
\centering
\caption{{\bf New York Times best-seller 
lifetime distribution parameters.} 
Listed are the linear and quadratic parameters 
$a$ and $b$, respectively, for the log-normal distribution 
Eq.~\eqref{p_L_H_max}, as shown in
Figure~\ref{fig_lifetime_distribution_nytimes}. Note 
that the prefactor of the linear term is $a+1$, see
(\ref{p_L_power_law}). The uncertainties 
refer to the margin of error given a 95\% confidence level.}
\label{tab_nytimes_lifetime_distribution_parameters}
\begin{tabular}{crr}
\\
\toprule
period & $a\!+\!1$ & $b$\\
\midrule
1959 -- 1968 & $1.81 \pm 0.70$ & $-0.49 \pm 0.44$\\
1969 -- 1978 & $2.02 \pm 0.88$ & $-0.69 \pm 0.55$\\
1979 -- 1988 & $1.15 \pm 0.72$ & $-0.12 \pm 0.46$\\
1989 -- 1998 & $0.06 \pm 0.71$ & $0.73 \pm 0.49$\\
1999 -- 2008 & $-0.33 \pm 0.66$ & $1.46 \pm 0.53$\\
2009 -- 2018 & $0.75 \pm 0.72$ & $0.85 \pm 0.50$\\
\bottomrule
\end{tabular}
\end{table}

The New York Times best-seller list and Billboard album charts 
are also similar with respect to the dynamics of titles that
eventually make it to the top, the number-one titles. For
this there are two possible courses. Either a title enters the
chart directly at the top, then its number-one position is
guaranteed. Alternatively a given title starts lower and
works its way up to the top, over the course of several
weeks or months. The probability for a number-one title 
to take the first course of action, $P_{\rm one}$, is shown in
Figure~\ref{nytimes_cd_pone}. In the 1970s and 1980s
close to no number-one best seller entered the New York Times 
best-seller list at the top. This changed dramatically in the 
following three to four decades. From the late 2000s onward 
close to all, about $95\%$ of all number-one novels started
charting right at the top. This is similar to the evolution 
the Billboard album charts took. A small difference is that 
the New York Times best-seller list started the transition 
about five years prior to the Billboard album chart, however 
with the evolution taking about ten years longer. In general 
the Billboard top 10 and top 100 charts differ only marginally.
 
The changes in $P_{\rm one}$ are also reflected in the time it 
takes on the average to become number one. Before 1980, when 
close to all number-one titles had to work their way 
up to the top, this process took on the average five to 
eight weeks. Today number-one novels as well as albums 
become number one in under a week on the average, since
almost all number-one titles start as such.

\begin{figure}[t]
\centerline{
\includegraphics[width=1.00\columnwidth]{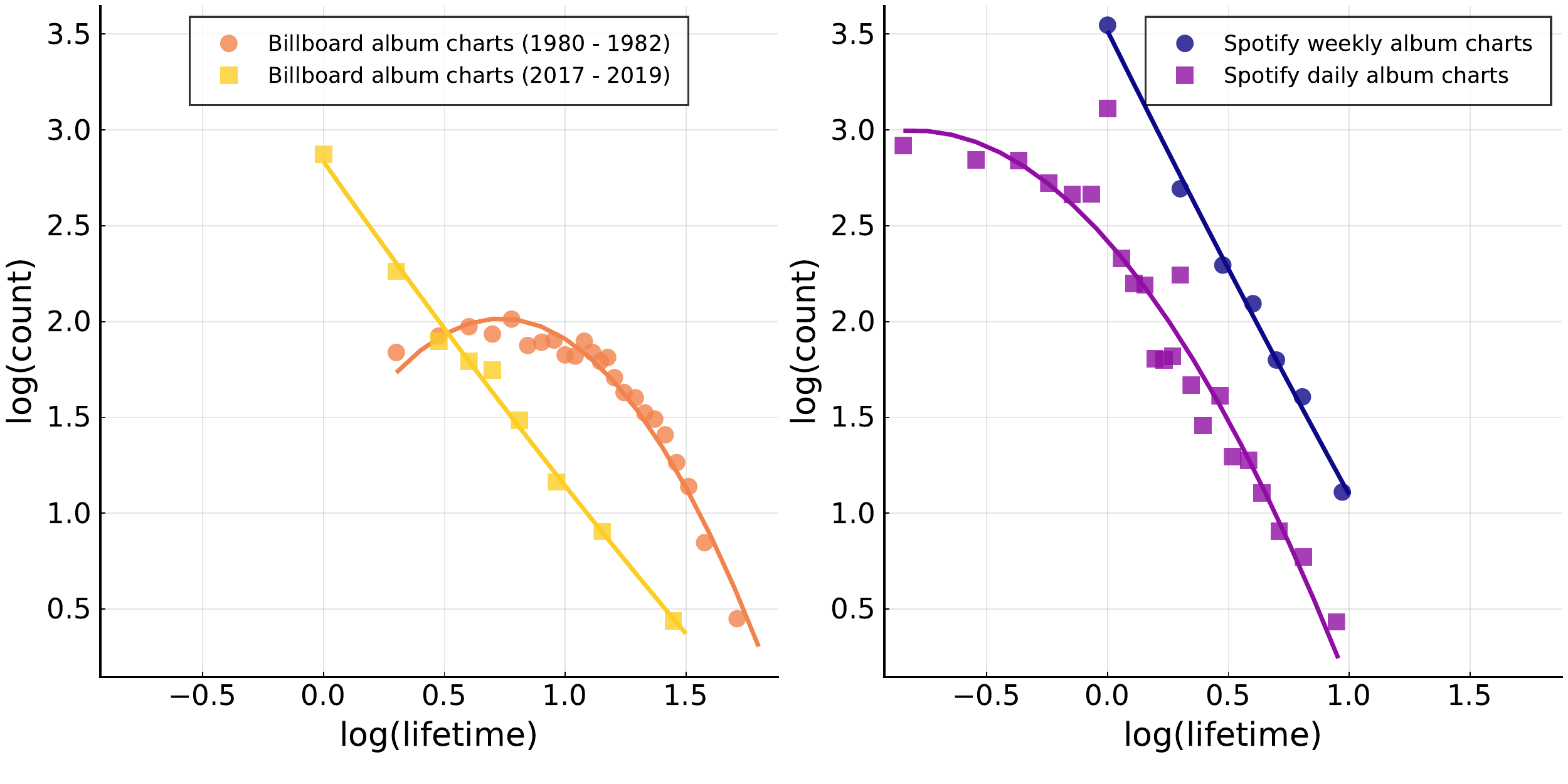}
           }
\caption{{\bf Lifetime distribution of music albums.} 
The probability of album lifetimes, the number of
weeks listed, for the Billboard (left) and Spotify 
(right) charts. A log-log representation has been used, for
which log-normal distributions correspond to parabolas
and power laws to straight lines, see Eq.~(\ref{p_s_H_max}).
Two three year periods, 1980-1982 (orange) and 2017-2019 (yellow)
are compared for the Billboard data. Daily (violet) and weekly 
(blue) Spotify album charts have been evaluated for 2017-2019.
Note the outlier for the Spotify daily charts  at a lifetime 
of exactly one week, which coincides to the minimal lifetime 
of the weekly chart. The Billboard charts are consistently 
compiled on a weekly basis.
}
\label{lifetime_distribution_music}
\end{figure}
 
\subsection{Musical chart lifetimes}

In Figure~\ref{lifetime_distribution_music} 
the lifetime distributions of daily and 
weekly music album charts are presented, namely
for the Billboard album charts, which are published
on a weekly basis, and for the respective
Spotify download charts. 

Comparing the functional dependency of past
(1980--1982) and present day (2017--2019) 
Billboard album lifetimes, one observes a transition
from a log-normal distribution to a power law,
as parameterized by \eqref{p_s_H_max}. An
equivalent transition is seen between daily 
and weekly Spotify charts. A caveat is here the
outlier that can be observed for the Spotify 
daily charts at a lifetime of exactly one week.
The origin of this outlier is unknown, it may 
be caused possibly by weekly algorithmic influences, 
such as maintenance periods. It is in any case
remarkable, that the lifetime distributions
presented in Figure~\ref{lifetime_distribution_music}
change on a functional level, either with time
or when the charting period is increased. We
argue, that a separation of time scales causes
both transitions.

\begin{figure}[t]
\centerline{
\includegraphics[width=1.00\columnwidth]{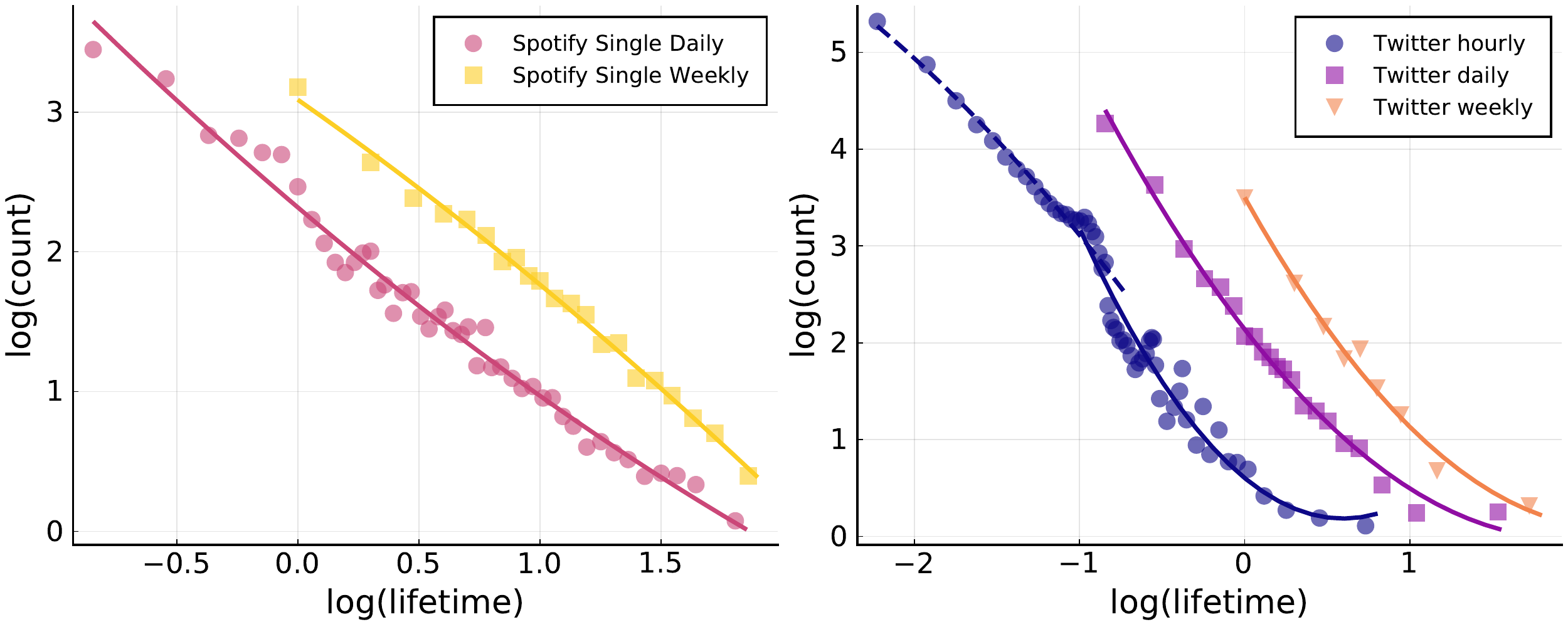}
           }
\caption{{\bf Spotify single downloads and Twitter hashtag popularity.} 
{\em Left:} Lifetime distribution of individual songs on Spotify 
daily and weekly single charts. Data for 2017--2019.
{\em Right:} For top-50 Twitter hourly, daily and weekly
hashtag charts the lifetime distribution. A kink 
corresponding to 24\,hours is present for the 1h charts.
Dashed/solid lines hourly charts correspond to quadratic
fits to the data below/above the kink.
Data from \citet{moensted2019}. 
}
\label{fig_SpotifySingle_TwitterHastags}
\end{figure}

Traditionally, music albums were bought in
music shops, which involved a personal trip,
and hence a comparatively long execution time.
Long decision and execution times imply, as 
argued in Sect.~\ref{sect_infoMax}, that the 
variance of the album lifetime distribution
is relevant, which takes consequently the form
of a log-normal distribution. The typical
time needed to acquire a music album dropped
however substantially below one week when 
online shopping started to become relevant 
in the 1990s. Given that one week corresponds
to the chart frequency, this development implies
that the variance of the album lifetime 
distribution lost its relevance. The desire 
to maximize information, the basis of the 
here discussed theory of statistical 
decision-making, leads in this case to power laws.

We believe that an analogous argument explains 
why daily and weekly Spotify album charts show
respectively log-normal and power law distributions.
On the average, it is presumably not a matter of 
only a few hours to listen and to appreciate music albums, 
which may contain a substantial number of titles, but
of a few days. This time scale, several days, lies
between the frequencies of daily and weekly
streaming charts, which would explain why daily
and weekly album charts are respectively 
log-normal and power law distributed.

Whereas the Billboard album charts are sales-based, 
the respective single charts are compiled
predominantly on the basis of airtime 
statistics, for which the number of times 
songs are played by radio stations are counted.
A direct comparison of album and single charts
is in this case not possible \citep{schneider2019five}. 
The equivalent caveat does not hold for the Spotify 
charts, for which both album and single charts are
based on streaming counts. 

In Figure~\ref{fig_SpotifySingle_TwitterHastags}
the lifetime distributions of daily and weekly
Spotify single charts are shown. Both are
close to power law distributions with only small
quadratic contributions. The overall process, to 
listen to a music sample, to decide to stream, 
and to actually do it, can be assumed to take 
only a few minutes for individual songs, but 
substantially longer in the case of albums.
The observation that already daily single 
download charts show indications for power law 
behavior, with respect to the distribution of 
song lifetimes, is hence consistent with our 
basic framework.

\subsection{New York Times book best sellers}

\begin{figure}[t]
\centerline{
\includegraphics[width=1.00\columnwidth]{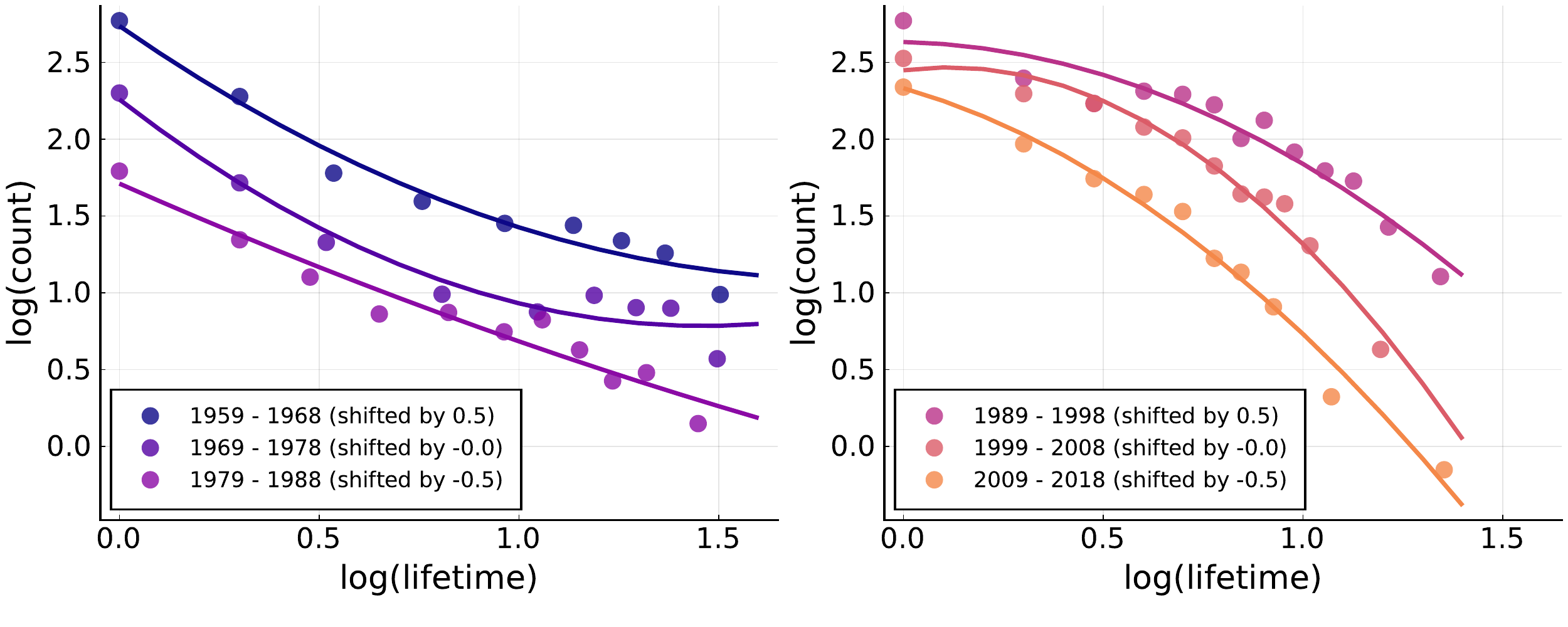}
           }
\caption{{\bf Six decades of NYT best seller lifetimes.} 
The lifetime distribution of books on the 
New York Times best-seller list. Shown is the  
average over ten-year periods. Note that 
only ten books are listed weekly, the data is 
hence comparatively sparse. In a log-log presentation,
as shown, a long-term evolution from convex
to concave is observed. 
}
\label{fig_lifetime_distribution_nytimes}
\end{figure}

In Figure~\ref{fig_lifetime_distribution_nytimes}
we present the probability that a given title stays 
for a certain number of weeks on the fiction best-seller
list of the New York Times. In contrast to the 
top-100 music charts discussed further above, only 
the top ten titles of the week are included in the 
NYT best-seller list. In order to compensate for the
comparatively limited database, we averaged the 
lifetime distribution over consecutive ten-year 
periods. Substantial scattering of the data can be
observed nevertheless. 

Included in Figure~\ref{fig_lifetime_distribution_nytimes}
are quadratic fits, $-(a+1)s-bs^2$, to the log-log 
representation of the lifetime distribution (compare 
Eq.~\eqref{p_s_H_max}). It is notable, that the 
distribution of book lifetimes evolves from being 
convex in log-log scale in the 50s and 60s to be
being concave, starting from the 80s. The turning
point, in the 70s, matches roughly a shallow
minimum in the chart diversity, as shown in
Figure~\ref{nytimes_cd_pone}. It is presently unclear
what drives this interesting phenomenon. The
presence of a finite second-order component, $b\ne0$,
indicates in any case that the individual time scales
for buying and reading books has not dropped below
one week, the charting frequency. A power law appears
in the 70s, when the lifetime distribution transits
from convex to concave. The power laws seen in the
distribution of musical charts emerge in contrast
in the concave region, see Figure~\ref{lifetime_distribution_music},
when an initial $b>0$ progressively vanishes
\citep{schneider2019five}. This indicates two 
distinct causalities for the two power laws, 
for the late-state Billboard lifetime 
distributions and for the book lifetimes in the 70s.

\subsection{Dynamics of Reddit comments}

New data items, like posts or images, are at the
core of social media activity. The response of
other users, in the form of comments or likes,
will determine the popularity of social media
posts and with this the likelihood of additional 
likes and comments. In this context we analyzed
the statistics of Reddit post commenting, using
publicly available data \citep{Reddit}. Note, 
that Reddit is a discussion website organized around 
collections of user-created discussion boards for 
individual subjects called ``subreddits''. Users 
can submit posts such as text posts, links, images
and videos to a subreddit. Posts are then voted up or 
down and commented on. While upvoting can be done 
directly on the post overview screen while scrolling 
through, in order to comment, users have to actively
click on the post.

The most popular posts from each subreddit are also shown 
on the front page on login. There are multiple options for sorting 
to determine which posts are the most popular. The old default 
was ``Hot'', which was based on the number of upvotes 
(log-weighted) and the submission time (recent posts 
prioritized). In 2018, ``Best'' was introduced as the new default,
which tries in addition to show new content, including weighting 
based on comment activity. Comment data is publicly 
available \citep{Reddit}, including precise timestamps, 
vote counts are available however only as monthly averages.
The technical aspects of our data analysis is described
in Sect.~\ref{sect_Reddit}.

For book and album charts the ranking criteria are the 
number of sales and downloads within the respective chart 
periods. In analogy, we used the number of comments
for Reddit posts. In Figure~\ref{reddit_lifetimes} the 
chart diversity and the respective lifetimes for 10- 
and 60-min Reddit top-100 charts are shown. Within the time 
window analyzed, 2013-2015, a downward trend in chart
diversity is observed, with an equivalent increase in
chart lifetimes. This is in contrast to the 
long-term trend for book and album charts that can
be observed in Figure~\ref{nytimes_cd_pone}.

The distributions for 10- and 60-min chart lifetimes
presented in Figure~\ref{reddit_lifetimes} is 
non-monotonic. On a coarse level, the probability
to observe a certain lifetime drops in a
power law-like fashion, which is however 
interseeded by a pronounced local maximum. 
The maxima correspond to characteristic time scales
of about 20-26 hours, for both the 10- and the 60-min
charts, which suggests that the intrinsic 24\,hours
day-night activity cycle may be involved. Users
may sleep on a post or comment of the day, in order
to revisit it in the next morning. A lesson learned
from the Reddit database is then, that the presence of
characteristic time scales may interfere strongly
with the otherwise operative feedback loop between 
posts and comments.

\subsection{Twitter Hashtags}

We use a publicly available corpus of
Twitter hashtag statistics to study the 
residence time of hashtags in hourly, daily 
and weekly top-50 charts. See Sect.~\ref{sect_Twitter}
for data source and handling. Hashtags and
Reddit comments are both examples of active 
user contributions, albeit with a key difference.
Adding a hashtag to a tweet can be presumed to
be on the average less time-consuming than 
composing an entire Reddit comment. We may hence
expect the trend towards self-organization to
be more pronounced for Twitter hashtag charts, 
than for Reddit comments. 

\begin{figure}[t]
\centerline{
\includegraphics[width=0.5\columnwidth]{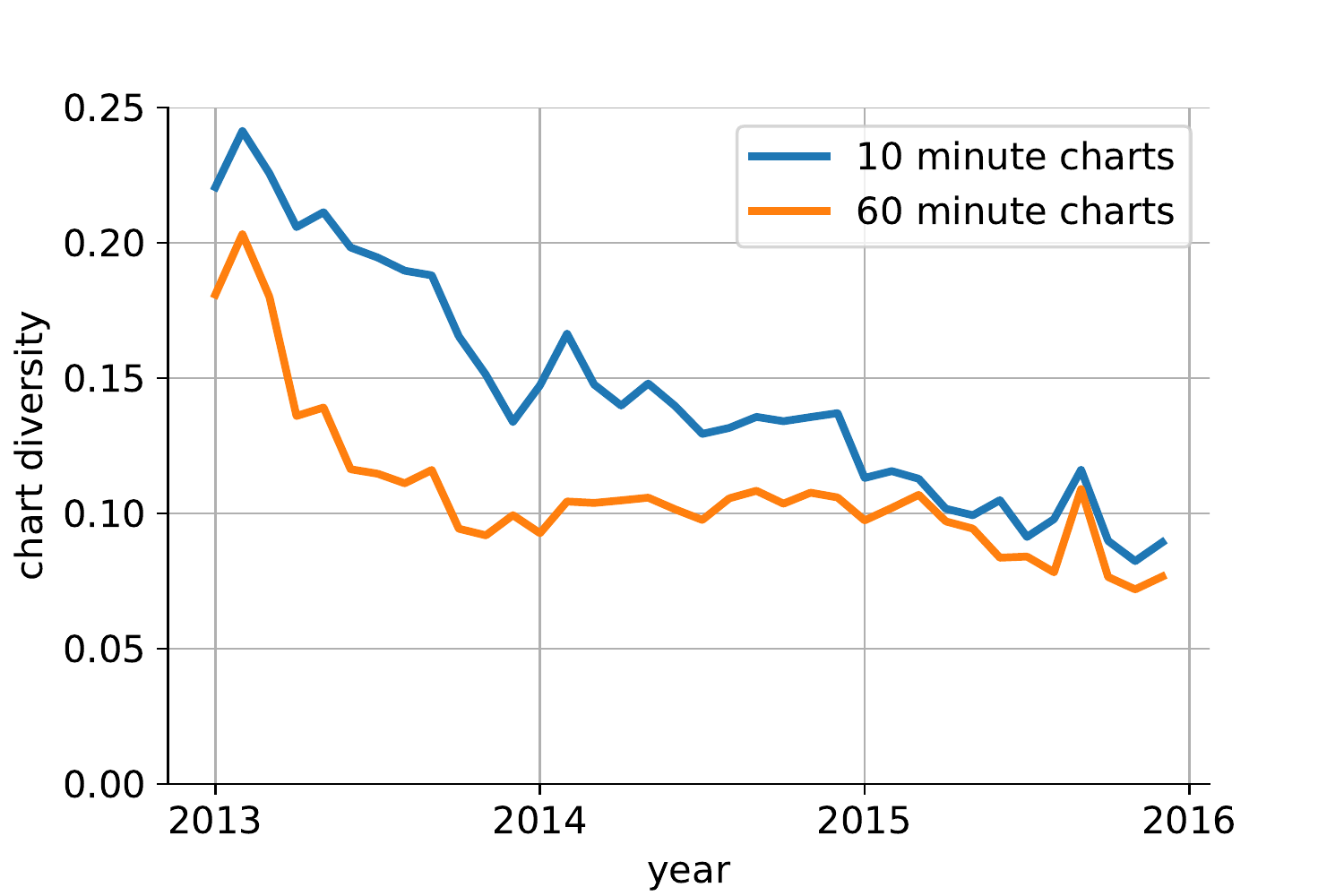}
\includegraphics[width=0.5\columnwidth]{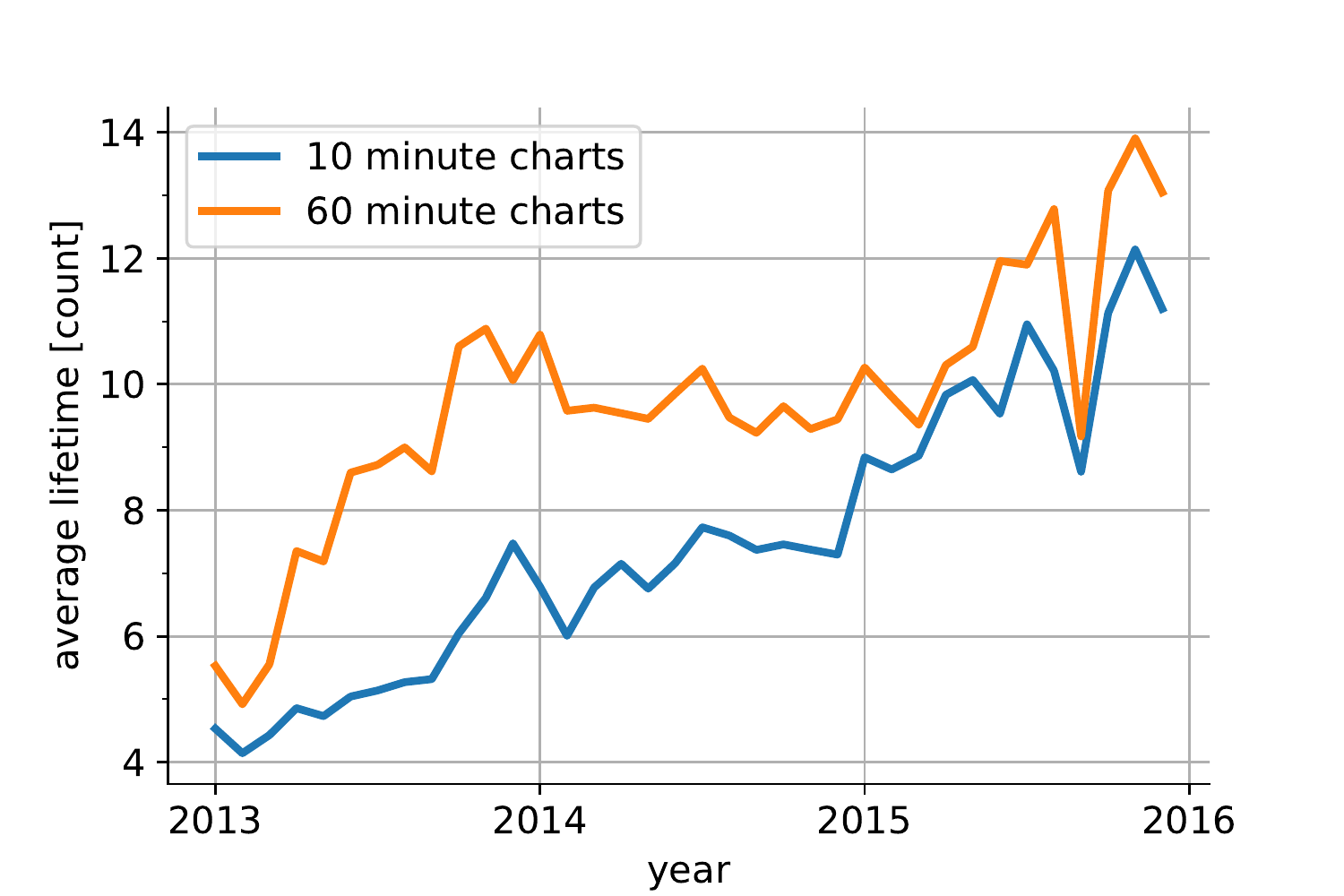}
           }
\centerline{
\includegraphics[width=0.5\columnwidth]{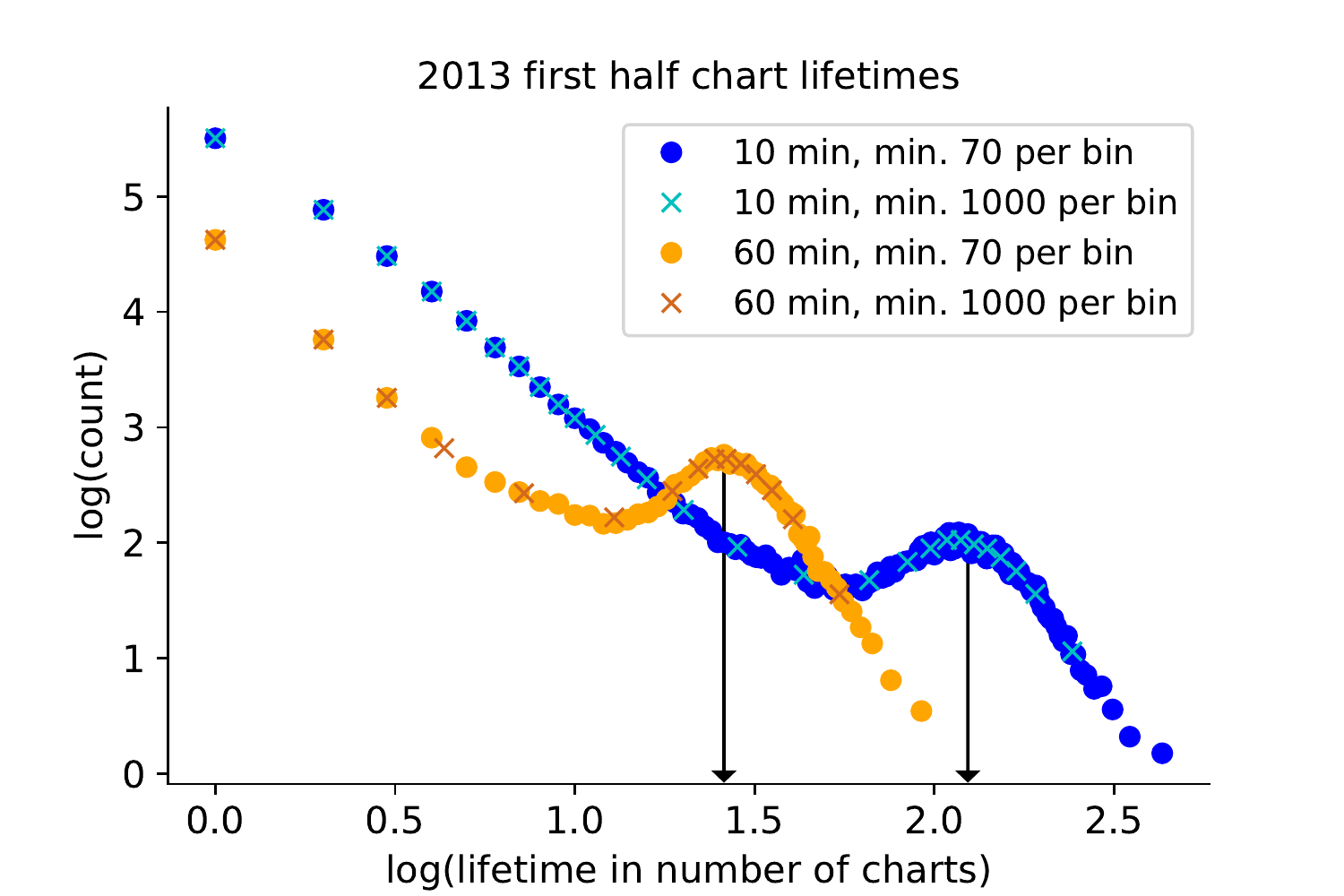}
\includegraphics[width=0.5\columnwidth]{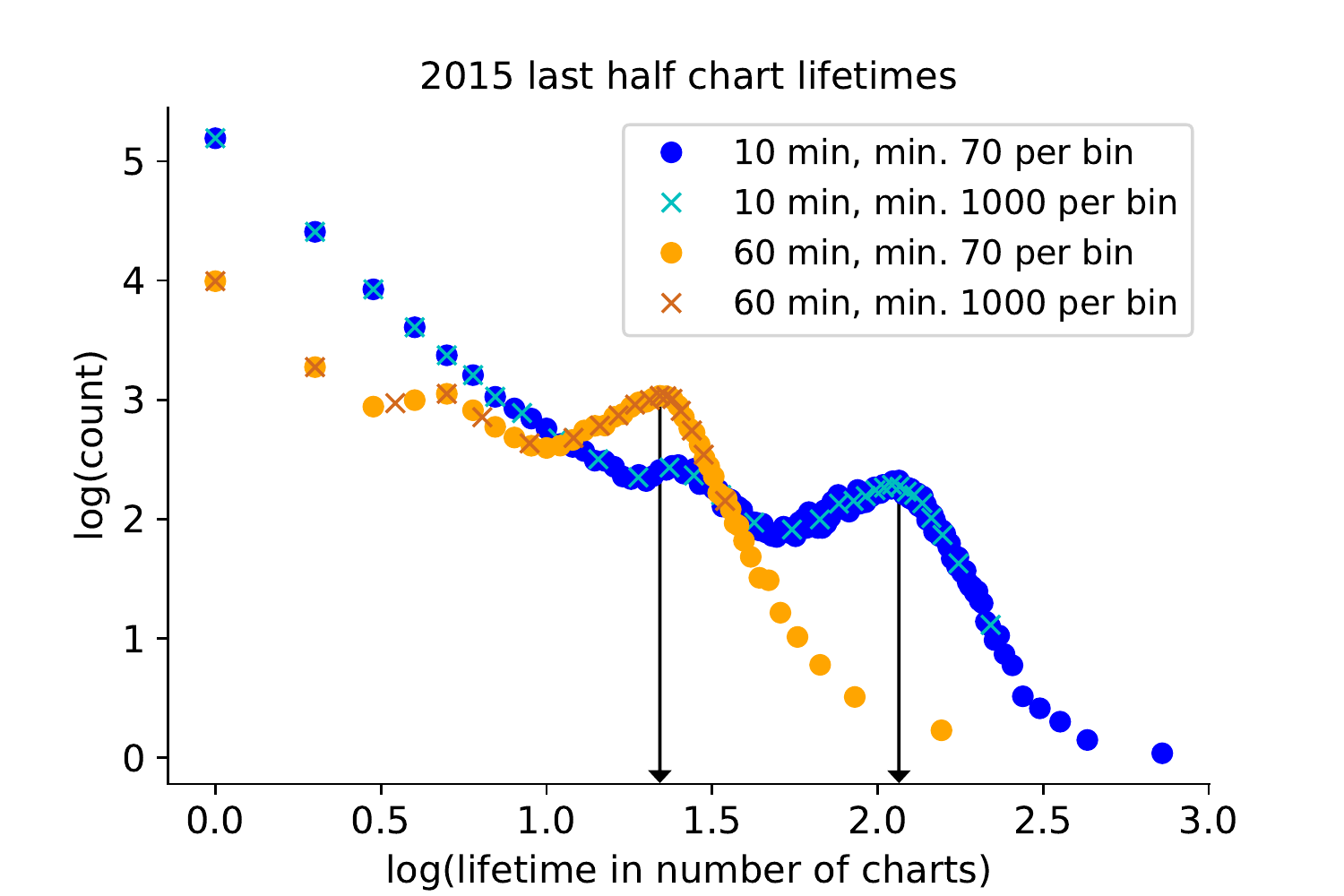}
           }
\caption{{\bf Reddit comment charts.} 
{\em Top:} The evolution of the diversity of 10/60\,min comment
charts over the observation period (left) and the respective
average lifetimes (right).
{\em Bottom:} Respectively for the first six months of 2013 and
the second half of 2015, the distributions of comment lifetimes,
both for 10min and 60\,min charts. Adaptive binning has been
used, as described in Sect.~\ref{sect_binning}, with either 
a minimum of 70 or 1000 comments per bin. The local maxima
in the lifetime distribution (vertical arrows) correspond
to 1240/1560 minutes (20.7/26.0 hours) for the 2013 
10/60\,min charts and to 1160/1320 minutes (19.4/22.0 hours)
for the 2015 10/60\,min charts.
See Sect.~\ref{sect_Reddit} for data processing.
}
\label{reddit_lifetimes}
\end{figure}

In Figure~\ref{fig_SpotifySingle_TwitterHastags}
the distribution of Twitter hashtags for
hourly, daily and weekly top-50 charts are presented.
With respect to the Reddit data shown in
Figure~\ref{reddit_lifetimes} one observes that 
the 24\,hour activity cycle is now substantially less 
pronounced. Instead of a peak, the dominant feature of
the lifetime distribution of the hourly Twitter hashtag 
charts is a kink. To be precise, only a weak local maximum 
occurs at about 18h for the hourly charts, and a 
30\% drop at 24h. For daily and 
weekly charts no anomaly is observed. Fitting the
Twitter hashtag data with the maximum entropy distribution
(\ref{p_s_H_max}), we decided to split the 1h lifetime 
data in two parts, below and above the above discussed
kink. Within the log-log representation, one finds that 
the quadratic contribution tends to become smaller for 
longer charting periods, viz when going from 1h to daily 
and weekly charts. This observation complies with the
argumentation laid out in Sect.~\ref{sect_infoMax},
namely that lifetime distributions become more
power law-like when the time scale of the individual
activities is substantially smaller than the charting
periods.

\section{Data sources and processing\label{sect_data}}

\subsection{Book and album charts\label{sect_charts}}

The New York Times book best sellers \citep{NYTbestseller},
the Spotify music streaming charts \citep{spotifyCharts}
and the Billboard sales-based music charts \citep{billboardCharts}
have been obtained from public internet sources.
We also examined the statistics of 
the set of Twitter hashtags
compiled by \cite{lorenz2019accelerating}.

The algorithm used for the compilation of the
Billboard charts has been adjusted over time, 
with a major update in 2014/15. At that point 
the traditional sales-based ranking was substituted 
by a ranking based on a multi-metric consumption 
rate, which includes weighted song streaming. 
This update, which took effect at the end of 2014, 
affects the chart statistics profoundly.

The original Billboard album chart, the Billboard Top 200,
was retained after the 2014/15 metric update under a new name,
as `Top Album Sales'. Data continuity is consequently achieved
when using the Top Album Sales charts from 2014/15 on. 
Whenever possible, we show results obtained from both the
Billboard Top 200 and the Top Album Sales charts, where the latter
are compiled according to the unaltered sales-based ranking rules,
albeit only for 100 ranks.

For data analysis we used statistically relevant binning,
which consists, as explained in Sect.~\ref{sect_binning}, 
of adjusting bin sizes dynamically such that a minimal 
number of $N_{\rm min}^{\rm (data)}$ data points per 
bin is obtained.

\subsection{New York Times Best-seller List\label{sect_nytimes}}

In contrast to the Billboard magazine, the 
New York Times does not publish a full
history of their best-seller list. Only 
recent lists are published on the official 
New York Times website \citep{NYTbestseller}. 
Consequently, the analysis presented here relies 
on a republication of the data by \citet{nytimesArchiv},
which collected and republished the weekly 
best-seller list from the 1950s until today. 
Since the length of the best-seller list changed 
repeatedly, all analysis above consider
a top 10 ranking.

\subsection{Reddit data analysis\label{sect_Reddit}}

Reddit data was downloaded from the \citet{Reddit}. Due
to the large data size, we restricted our analysis
to two six-month periods, the first half of 2013,
which contains 19.050.122 post and 187.385.130 comments, 
and the second half of 2015, with a total of 
37.038.895/347.236.797 entries. Comments of posts 
receiving less than 1000 comments over the observation 
period were removed from the data, since they would 
not show up in the charts analyzed. The cleaned 
comment data was separated into time slices 
based on the comment timestamps and aggregated 
by post IDs (identification labels). In this way
charts for consecutive 10 minute and 60 minute slices
where generated, as based on the number of comments
received. Post lifetimes were evaluated from the
respective top-100 charts.

\subsection{Twitter Hashtags\label{sect_Twitter}}

The hourly ranking of the 50 most common hashtags were gathered 
by \citet{moensted2019}. This dataset, which has been used to
study the acceleration of collective attention
\citep{lorenz2019accelerating}, contains the 50 most common 
hashtags out of a 10\% sample of all tweets, which were gathered 
every hour during the period from 2013 through 2016. For every 
hashtag the dataset also shows the number of times it appeared 
during the last hour in the 10\% sample. This additional information 
allows, modulo the top-50 cut-off, to compile charts with longer 
observation intervals, e.g.\ daily or weekly charts.

\subsection{Adaptive binning\label{sect_binning}}

For the generation of probability distribution
functions from collected data one needs to
group events together into bins. When sets of bins
with predefined width are used, the number of
events may vary strongly from one bin to another.
It is hence advantageous, for statistical relevant 
binning, to adjust the width of the individual
bins dynamically, until a minimum of
$N_{\rm min}^{\rm (data)}$ data points per bin
has been reached. The binning procedure is finished 
once this is not any more possible. Comparing
results obtained for different $N_{\rm min}^{\rm (data)}$,
as done in Figure~\ref{reddit_lifetimes}, allows
to gauge the accuracy.

\section{Discussion}

Consumption charts, like sales and streaming charts, show 
remarkable similarities between different cultural products, 
e.g.\ when comparing music albums and literature. The evolution 
of the respective chart diversities, as well as the probability 
that a number-one title debuts as such, follow parallel 
courses for these two cultural goods. Similarly, one can fit 
the distribution of on-chart residence times, the respective
lifetimes, in all cases by a generalized log-normal
distribution that can be shown to maximize the 
information content of the lifetime distribution, to
the extent as it is stored in the brain. Taking the long-term 
perspective, one notices distinct features for the evolution 
of the lifetime distribution of the New York Times best-seller 
list, and for album sales charts. 

For music albums \citep{schneider2019five}, the 
lifetime distribution evolves from a log-normal 
distribution, as before the 1990s, to a power law,
as for nowadays. In this context it is important to notice
that power laws belong to the class of generalized log-normal
distributions, albeit with a vanishing quadratic term in
the exponent. The information-theoretical arguments
presented in this study for the occurrence of log-normal
distributions suggests that a transition from fully 
log-normal to a power law takes place when uncertainties,
viz the variance, fade into secondary importance. This
phenomenon is related to the time needed on the average
for decision-making, given that it takes longer to gauge
variances than means.

In societies, some time scales remain constant, others
are subject to a secular acceleration process
\citep{rosa2013social}. An example of a non-changing time 
scale is the period of typically 4-5 years between general 
elections, which contrasts with an accelerating opinion
dynamics \citep{gros2017entrenched}. Of relevance for 
the present study is the charting period of one week, 
which did not change since the inception of classical 
music charts and book bestseller lists. In contrast to 
the charting period, the decision time to buy an album
is likely to have fallen substantially with the rise of 
the internet. Here we have argued that this development 
is reflected in the respective chart statistics. It is 
less clear if the same holds for the time people need 
to read a book.

Modern streaming charts can be used as a test bed for the
relative time scale hypothesis. The available high
temporal resolution allows narrowing down the time 
individuals need to decide to listen to an album, and 
then to actually do so. The lifetime distributions of 
Spotify album charts show that this time scale is 
nowadays between one week and one day. For a single song 
the decision can be made comparatively faster, which
leads for both weekly and daily single charts to power law 
lifetime distributions. 

Activity charts, such as Twitter hashtags or Reddit postings, 
differ in certain aspects from consumption charts. Human
activities are subject to a day-night cycle, which shows up
for the case of commenting on Reddit as a resurgence after around 
24\,hours. Previous-day posts are likely to be revisited after 
a night of sleep, as evident in Figure~\ref{reddit_lifetimes}.
This phenomenon, the 24h-cycle, will mask underlying 
power laws, if existent. The Twitter charts provide evidence 
for an equivalent occlusion mechanism, see
Figure~\ref{fig_SpotifySingle_TwitterHastags}. 

The lifetime distribution of hourly hashtag charts 
is characterized by a peak at 18\,hours, together 
with a subsequent drop at 24\,hours. Charts
averaging over one or more full days, daily and 
weekly charts, are represented instead by smooth 
distributions. One finds indications for power laws 
with minor quadratic corrections. These results 
suggest that power laws appear ubiquitously 
for long-enough charting periods. In conclusion 
we believe that the data examined supports the 
basic hypothesis presented here, namely that 
aggregated human decision processes exhibit the 
pronounced statistical features characteristics 
of compressed information maximization.

Beyond information maximization, we found 
marked changes at the top of cultural charts,
which started with the raise of the internet
in the 1990s. This phenomenon regards the 
route to become a number-one hit, which took
substantially longer in the past. We believe
that this observation deserves further investigations,
which would however transcend our present 
framework.

\section*{Acknowledgement}

The work of BS was supported by the research grant of the 
Romanian National Authority for Scientific Research and Innovation, 
CNCS/CCCDI-UEFISCDI, project nr.\ PN-III-P4-ID-PCCF-2016-0084.

\section*{Data availability}

All data is available from publicly 
accessible sources, as described in
Sect.~\ref{sect_data}.



\end{document}